\documentclass[a4paper,11pt]{article}
\usepackage{pos}
\usepackage{slashed}
\usepackage{multirow}
\usepackage{subcaption}
\usepackage{hyperref}
\usepackage{subcaption} 
\usepackage{setspace}

\let\OLDthebibliography\thebibliography
\renewcommand\thebibliography[1]{
  \OLDthebibliography{#1}
  \setlength{\parskip}{0pt}
  \setlength{\itemsep}{0pt plus 0.3ex}
}

\newcommand{\be}{\begin{equation}}
\newcommand{\ee}{\end{equation}}
\newcommand{\ba}{\begin{eqnarray}}
\newcommand{\ea}{\end{eqnarray}}

\graphicspath{{./Figures/}}

\newcommand{\eref}[1]{(\ref{#1})}

\title{
Coordinate-space calculation of QED corrections to the\\ hadronic vacuum polarization contribution to $(g-2)_\mu$ }

\author[d]{En-Hung Chao}
\author[a,b,c]{Harvey B. Meyer}
\author*[a]{Julian Parrino}

\affiliation[a]{PRISMA$^+$ Cluster of Excellence \& Institut f\"ur Kernphysik,\\
Johannes Gutenberg-Universit\"at Mainz,
D-55099 Mainz, Germany}

\affiliation[b]{Helmholtz~Institut~Mainz,\\
Staudingerweg 18, D-55128 Mainz, Germany}

\affiliation[c]{GSI Helmholtzzentrum f\"ur Schwerionenforschung, \\ D-64291 Darmstadt, Germany}

\affiliation[d]{Physics Department,
Columbia University,
New York, New York 10027, USA}

\emailAdd{juparrin@uni-mainz.de}

\abstract{As several lattice collaborations agree on the result for the
 window quantity of the hadronic vacuum polarization (HVP) contribution to
 $(g-2)_\mu$, whilst being in tension with the calculation using the dispersive
 approach, further effort is needed in order to pin down the cause for this
 difference.
Here we want to focus on the isospin breaking corrections to the leading order HVP.
 In many lattice applications, the photon propagator is treated
stochastically; however, by analogy with the hadronic light-by-light
contribution (HLbL) to  $(g-2)_\mu$,
 we suggest a coordinate-space approach to the HVP at next-to-leading order. We
 present a calculation of the two diagrams of the (2+2) topology at
 unphysical pion mass, where we apply a Pauli-Villars regularization for the extra
 photon propagator in the diagram that is UV-divergent. We compare the
 UV-finite diagram to the pseudoscalar exchange contributions calculated from a vector-meson dominance model.}

\FullConference{%
The 40th International Symposium on Lattice Field Theory (Lattice 2023)\\
July 31st - August 4th, 2023\\
Fermi National Accelerator Laboratory
}

\begin{document}\let\oldref\ref

\addtocounter{page}{-1}
\maketitle

\section{Introduction}
The anomalous magnetic moment of the muon $a_\mu$ has been in the focus of the particle physics community for many years now. Recently the experimental result reached an uncertainty at the 0.20 ppm level, see Ref.~\cite{Muong-2:2023cdq}. This result is in $5.0 \sigma$ tension with the theory result given in the 2020 White Paper, Ref.~\cite{Aoyama:2020ynm}. The uncertainty of this theory value is entirely dominated by the hadronic contributions, with the hadronic vacuum polarization (HVP) making the largest contribution. There are two different approaches to calculating of the HVP  contribution $a_\mu^{HVP}$: the dispersive approach  and lattice QCD.
Recently it has been shown that there is a significant tension between the different methods when calculating the intermediate window quantity $a_\mu^W$, a partial contribution to the total $a_\mu^{HVP}$ that is easier to calculate on the lattice. Another problem came up with the new measurement at CMD-3, see Ref.~\cite{CMD-3:2023alj}: the measured $(e^+ e^- \rightarrow \pi^+ \pi^-)$ cross-section, which is the single most important data input for the dispersive approach, is not consistent with older measurements. These tensions have to be resolved, before a combined theory value of $a_\mu^{HVP}$ can be quoted.

On the other hand, there is a full lattice QCD calculation of $a_\mu^{HVP}$ claiming sub-percent precision, see Ref.~\cite{Borsanyi:2020mff}. It is important to have independent checks of this calculation in order to fully resolve the $(g-2)_\mu$ puzzle. In these calculations, QCD is treated non-perturbatively, but the different contributions can be expanded in the fine-structure constant $\alpha$. The leading contribution to $a_\mu^{HVP}$ is of order $O(\alpha^2)$. For a calculation of the HVP at sub-percent precision it is necessary to investigate $O(\alpha^3)$ corrections as well. These corrections to $a_\mu^{HVP}$ involve the same hadronic four-point function as in the calculation of the hadronic light-by-light (Hlbl) contribution $a_\mu^{Hlbl}$, see Refs.~\cite{Chao:2021tvp,Blum:2023vlm}, where QED is treated in the continuum and only the QCD four-point function is calculated on the lattice.
In this work, we propose a similar approach using a coordinate-space method for calculating the HVP contribution at next-to leading order $a_\mu^{HVP,NLO}$. We explain the basic formalism in Sect.~\ref{sect:ccs}. In the scope of these proceedings we will not cover all the diagrams contributing to $a_\mu^{HVP,NLO}$ at $O(\alpha^3)$, but focus on the $(2+2)$ disconnected contribution. By analogy with the Hlbl contribution, we expect that the $(2+2)$ together with the fully-connected contribution give the dominant part, when the photon propagator is regulated on hadronic distance scales. The method for calculating these diagrams is given in Sect.~\ref{sect:setup}. We then use a model, given in Sect.~\ref{sect:pme} to describe the integrand of the UV-finite $(2+2)$ diagram and compare the relative size of both $(2+2)$ contributions, see Sect.~\ref{sect:results}.
\section{Covariant coordinate-space method}
\label{sect:ccs}
The covariant coordinate-space (CCS) method for evaluating $a_\mu$, first derived in Ref.~\cite{Meyer:2017hjv} has been successfully tested to reproduce the same result as the time-momentum representation (TMR) for the intermediate window quantity at a pion mass of $\sim 350$ MeV in Ref.~\cite{Chao:2022ycy}. By expanding the QCD path integral to next-to-leading order in the electromagnetic coupling, one is able to express the HVP at NLO in the CCS representation, as shown in Ref. \cite{Biloshytskyi:2022ets}. In contrast to the Hlbl contribution, the HVP at NLO is UV-divergent. To handle this divergence, we use a Pauli-Villars regulated photon propagator, see Eq.~\eref{eq:pv}, with photon mass $\Lambda$.
In Feynman gauge we obtain
\ba
    \label{hvpkernelintegral}
    a^{HVP,NLO}_\mu=-\frac{e^2}{2}\int_{x,y,z}H_{\mu\sigma}(z)\delta_{\nu \rho}\Big[G_0(y-x)\Big]_\Lambda \widetilde{\Pi}_{\mu \nu \rho \sigma}(x,y,z) 
\ea 
with 
\vspace{-0.8cm}
\ba
\label{eq:pv}
\Big[G_0(y-x)\Big]_\Lambda=\frac{1}{4\pi^2|y-x|^2}-\frac{\Lambda K_1(\Lambda|y-x|)}{4\pi^2|y-x|},
\ea
the QCD four-point function $\widetilde{\Pi}_{\mu \nu \rho \sigma}(x,y,z)=\langle j_\mu(z) j_\nu(y) j_\rho(x) j_\sigma(0) \rangle_{QCD}$, the CCS kernel 
$H_{\mu \sigma}(z) = -\delta_{\mu\sigma}\mathcal{H}_1(|z|) +\frac{z_\mu z_\sigma}{|z|^2}\mathcal{H}_2(|z|)$ and the modified Bessel function of the second kind $K_1(x)$.
The QCD vacuum expectation value $\langle \dots \rangle_{QCD}$ implies that gluon interactions between all valence quark lines are taken into account.
Since the electromagnetic vector current $j_\mu(z)$ is conserved, we can add a total derivative to the kernel without changing the continuum result in infinite volume. Thus we can define the 'TL' (traceless) and 'XX' kernel
\be
\label{eq:kernels}
H^{\textrm{TL}}_{\mu\nu}(z) = \left(-\delta_{\mu \nu} +4\frac{z_\mu z_\nu}{|z|^2}\right)\mathcal{H}_2(|z|)\,, \quad
H^{\textrm{XX}}_{\mu\nu}(z) = \frac{z_\mu z_\nu}{|z|^2}\Big(\mathcal{H}_2(|z|) +|z|\frac{d}{d|z|}\mathcal{H}_1(|z|) \Big).
\ee
Evaluating Eq.~\eref{hvpkernelintegral} for both kernels serves as a consistency check. 
Analogously to Eq.~\eref{hvpkernelintegral} one can write down an expression for the window quantity $a_\mu^W$ and the subtracted vacuum polarization $\hat{\Pi}(Q^2)$ in the CCS representation, where only the weight functions $\mathcal{H}_1$ and $\mathcal{H}_2$ need to be changed. These functions are given in Ref.~\cite{Meyer:2017hjv} for $a_\mu$ and  $\hat{\Pi}(Q^2)$. The weight functions for the window quantity are derived in ~\cite{Chao:2022ycy}.\\
\section{Computing the (2+2) disconnected contributions}
\label{sect:setup}

\begin{figure}
    \centering
    \includegraphics[width=0.6\textwidth]{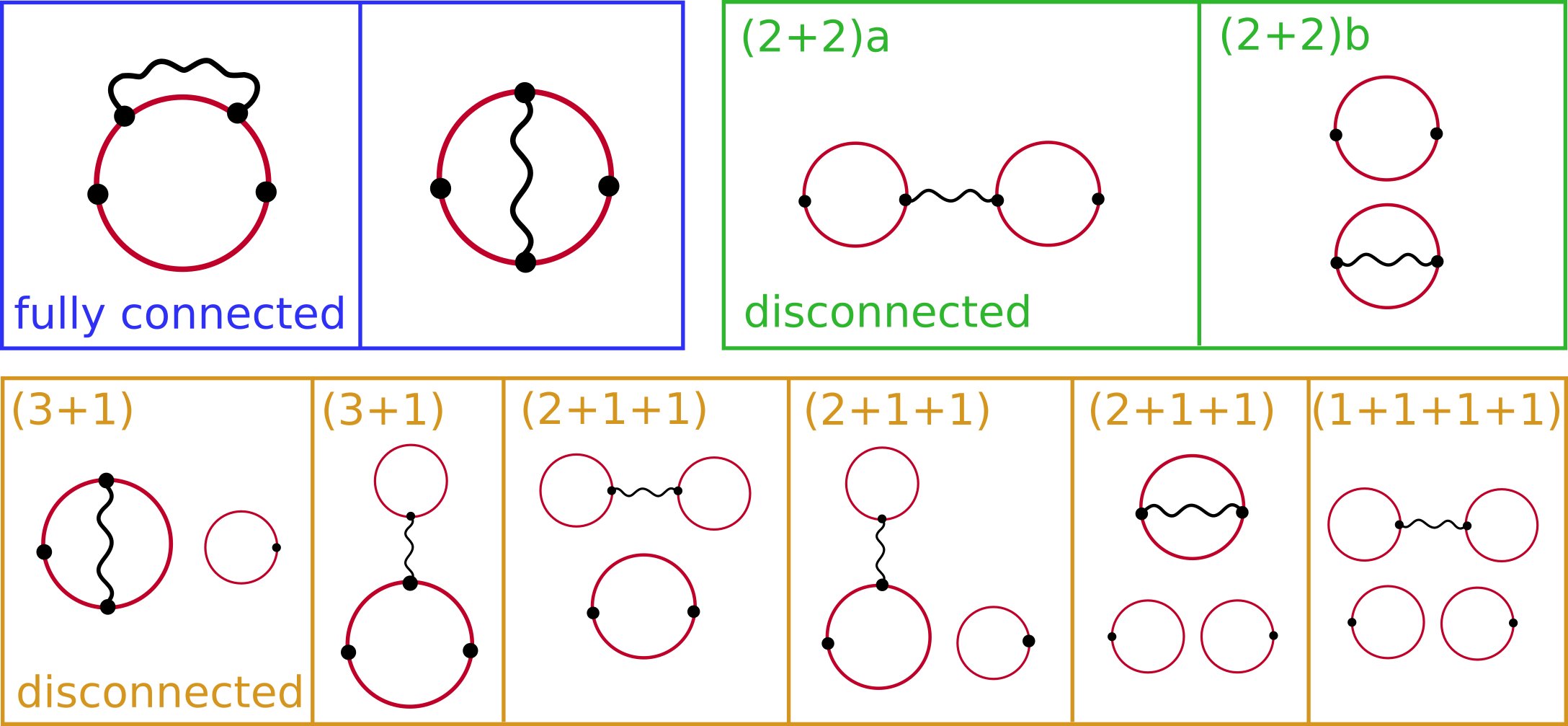}
    \caption{Diagrams that contribute to $\langle j_\mu(z) j_\nu(y) j_\rho(x) j_\sigma(0) \rangle_{QCD}$.  We categorize the relevant diagrams according to the number of vector currents (black dots) that are connected by valence quark propagators (red-lines).}
    \label{fig:diagrams}
\end{figure}

In order to calculate Eq.~\eref{hvpkernelintegral} it is important to inspect the different diagrams that contribute to the QCD four-point function. These are depicted in Fig.~\eref{fig:diagrams}. By analogy with the Hlbl contribution, we expect the fully-connected and the (2+2) disconnected to give the dominant part of $a_\mu^{HVP,NLO}$. Furthermore we note that all diagrams containing a self-contracted propagator are suppressed by the difference between the strange and the light quark masses due to the property $\sum_{f=u,d,s}Q_f=0$. In these proceedings we will only focus on the $(2+2)$ disconnected contributions with mass degenerate up and down (light) masses.
Performing the Wick-contractions, the four-point functions for the $(2+2)a$ and $(2+2)b$ read
\ba 
\label{eq:2+2a}
\widetilde{\Pi}^{(2+2)a}_{\mu \nu \rho \sigma}(x,y,z)=C \hat{Z}_V^4 2 \langle \hat{\Pi}_{\mu \nu}(z,x)\hat{\Pi}_{\rho \sigma}(y,0)\rangle_{G},
\ea 
\vspace{-0.9cm}
\ba 
\label{eq:2+2b}
\widetilde{\Pi}^{(2+2)b}_{\mu \nu \rho \sigma}(x,y,z)=C \hat{Z}_V^4 \langle \hat{\Pi}_{\mu \sigma}(z,0)\hat{\Pi}_{\rho \nu}(y,x)\rangle_{G}, 
\ea 
where the charge factor $C$ is given by $\frac{25}{81}$ for the light contribution, $ \hat{Z}_V$ is the renormalization factor for the electromagnetic vector current and $\langle \dots \rangle_{G}$ denotes the average over gauge configurations. In this notation the photon propagator connects the $x$ and the $y$ vertex and $z$ is the argument of the CCS kernel, see Eq.~\eref{hvpkernelintegral}.
The contributions are expressed through the two-point correlation function
\ba 
\label{eq:2ptf}
\Pi_{\mu \nu}(x,y) = -Re\Big(Tr\Big[S(y,x)\gamma_\mu S(x,y) \gamma_\nu  \Big] \Big),
\ea 
where the vacuum expectation value needs to be subtracted in order to avoid double counting of the contribution where the two QCD 'blobs' are not interconnected,
\ba 
\hat{\Pi}_{\mu \nu}(x,y) = \Pi_{\mu \nu}(x,y)-\langle \Pi_{\mu \nu}(x,y) \rangle_{G}.
\ea
Inserting Eqs.~\eref{eq:2+2a} and~\eref{eq:2+2b} into Eq.~\eref{hvpkernelintegral}. the integrals over $y$ and $z$ factorize and the final integral over $x$ only depends on its norm $|x|$. In the continuum and infinite-volume limit the $(2+2)$ contributions to $a_\mu^{HVP,NLO}$ now take the form 
\ba 
\label{eq:2+2_master_a}
a_\mu^{(2+2)a}=-\frac{e^2}{2}C Z_V^4 2\pi^2 2\int_0^\infty d|x| |x|^3 \Big[\langle I^{(2)}_{\rho \sigma}(x)I^{(3)}_{\sigma \rho}(x)\rangle_{G} -\langle I^{(2)}_{\rho \sigma}(x)\rangle_{G}\langle I^{(3)}_{\sigma \rho}(x)\rangle_{G} \Big],
\ea 
\vspace{-0.7cm}
\ba 
\label{eq:2+2_master_b}
a_\mu^{(2+2)b}=-\frac{e^2}{2}C Z_V^4 2\pi^2 \int_0^\infty d|x| |x|^3 \Big[\langle I^{(1)}(x)I^{(4)}\rangle_{G} -\langle I^{(1)}(x)\rangle_{G}\langle I^{(4)}\rangle_{G} \Big],
\ea 
with the four one-dimensional integrals
\ba 
\label{eq:I1}
I^{(1)}(x) = \int_z \Big[G_0(x-z)\Big]_\Lambda \Pi_{\nu \nu}(x,z),\quad I^{(2)}_{\rho \sigma}(x) = \int_y \Big[G_0(x-y)\Big]_\Lambda \Pi_{\rho \sigma}(y,0)
\ea 
\vspace{-0.7cm}
\ba 
I^{(3)}_{\sigma \rho}(x) = \int_z H_{\nu \sigma}(z) \Pi_{\nu \rho}(z,x), \quad I^{(4)} = \int_y  H_{\mu \sigma}(y) \Pi_{\mu \sigma}(y,0).
\ea
We use two-point correlation functions calculated on one ensemble generated by the CLS consortium with parameters given in Table~\eref{table:ensemble}. The simulation is performed with $N_f = 2 + 1 $ dynamical flavors of non-perturbatively $O(a)$ improved Wilson quarks and tree-level $O(a^2)$ improved Lüscher-Weisz gauge action.
Eq.~\eref{eq:2ptf} is calculated and stored for 24 different source positions at $y$ for all points on the lattice $x$. The point sources are distributed at $(2n,2n,2n,64)$ for $n=\{0,1,\dots,23\}$.
One of the sources is chosen as the origin and the integrands of Eqs.~\eref{eq:2+2_master_a} and~\eref{eq:2+2_master_b} are sampled over multiple values of $|x|$, while $x$ is given by the difference between two source positions. Using translational invariance on the lattice, we repeat this procedure choosing each of the source positions as the origin and averaging over the results for the same $|x|$ to increase statistics.

\begin{table}[]
\begin{tabular}{|c|c|c|c|c|c|c|c|c|}
\hline
Id   & $\beta$ & $L^3 \times T$ & $a$ [fm] & $m_\pi$ [MeV] & $m_K$ [MeV] & $m_\pi L$ & $L$ [fm] & \begin{tabular}[c]{@{}l@{}}\#confs \\ light\end{tabular} \\ \hline
N203 & 3.55    & $48^3 \times 128$             & 0.06426    & 346(4)            & 442(5)          & 5.4       & 3.1          & 180                                                          \\ \hline
\hline
\end{tabular}
	\caption{The parameters of the ensemble N203 generated by the CLS consortium. The lattice spacing is determined in Ref.~\cite{Bruno:2016plf} and the pion and kaon mass are taken from Ref.~\cite{Ce:2022kxy}.}

\label{table:ensemble}
\end{table}

\section{Comparison to the pseudoscalar meson exchange model}
\label{sect:pme}
To get a better understanding of the integrand, we employ a model calculation in infinite volume. Analogous to the the hadronic light-by-light scattering, see Ref.~\cite{Chao:2020kwq},  we expect the dominant part of the $(2+2)$ contributions to be explained by the pseudoscalar meson exchange (PME)
\ba
\nonumber
a_\mu^{{\rm hvp},PME} =  -\frac{e^2}{2} \int d^4xd^4yd^4z H_{\sigma\lambda}(z) \Big[ G_0(x-y)\Big]_\Lambda \int\frac{d^4qd^4kd^4p}{(2\pi)^{12}} e^{i(p\cdot z + q\cdot x + k\cdot y)} \Pi_{\sigma\mu\mu\lambda}(p,q,k)
\ea
\vspace{-1.1cm}
\ba
\label{eq:pi0master}
\ea
where the momentum-space four-point function in Euclidean spacetime, taken from Ref.~\cite{Knecht:2001qf}, is given by 
\vspace{-0.5cm}
\ba
\label{eq:pi_momentun}
\Pi_{\sigma\mu\mu\lambda}(p,q,k) &=&
\epsilon_{\sigma\mu\alpha\beta} \epsilon_{\mu\lambda\gamma\delta} \,p_\alpha
\Big(\frac{{\cal F}(-p^2,-k^2)\;{\cal F}(-q^2,-(p+k+q)^2)}{(p+k)^2+m^2}\; k_\beta\, q_\gamma(p+k)_\delta
\nonumber\\ && + \frac{{\cal F}(-p^2,-q^2) \,{\cal F}(-k^2,-(p+k+q)^2)}{(p+q)^2 + m^2}\; q_\beta\,k_\gamma (p+q)_\delta\Big).
\ea
We use the vector-meson dominance (VMD) parametrization of the transition form factor, see Ref.~\cite{Knecht:2001qf}
\be
{\cal F}(-p^2,-k^2) = \frac{{\cal F}(0,0)m_V^4}{(p^2+m_V^2) (k^2+m_V^2)}.
\ee
This allows us to simplify Eq.~\eref{eq:pi0master} such that we get an integrand, that only depends on the absolute value $|x|$ in the same fashion as Eq.~\eref{eq:2+2_master_a}.

The PME does not contribute to the $(2+2)b$ diagram. This can be seen from the fact that the total $(2+2)$ contribution is proportional to the t-channel pseudoscalar exchange, as worked out in the appendix of Ref.~\cite{Chao:2020kwq}. In the $(2+2)b$ diagram the incoming and outgoing momenta of the same vertex are contracted and thus give zero.
For the $(2+2)a$ diagram the $\pi^0$ contributes with a chargefactor of $-\frac{25}{9}$ while the $\eta$ and $\eta'$ contribute with factor $1$. 
For each of the pseudoscalar mesons we have 3 parameters in the model $(m,m_V,f)$, where ${\cal F}(0,0)=(4\pi^2 f)^{-1}$. For the pion we have $m=m_\pi$, $m_V=m_\rho$ and $f=f_\pi$ in the chiral limit. However, in order to compare the model to the lattice data we need to choose the parameters to match the corresponding values on the specific ensemble we use, see Table~\eref{table:ensemble}. The pion mass $m_\pi$ and decay constant $f_\pi$ are taken from Ref.~\cite{Ce:2022kxy}. For the $\rho$-meson mass $m_\rho$ we use the results of a VMD fit to the data from Ref.~\cite{gerardin:2019rua}, where the fit is restricted to the single virtual case, which was done in Ref.~\cite{Chao:2020kwq}.
We approximate the $\eta$ mass using the Gell-Mann--Okubo formula $m_\eta^2 \sim \frac{4}{3}m_K^2-\frac{1}{3}m_\pi^2$. The parameter $f_\eta$ is estimated by determining the value for ${\cal F}_\eta(0,0)$ using a linear interpolation in $\xi=m_K^2-m_\pi^2$ between its value at the physical point ${\cal F}^{\textrm{phys}}_\eta(0,0) \sim 0.27 \textrm{ (GeV)}^{-1}$ and the $SU(3)$-flavor symmetric point ${\cal F}^{\textrm{SU(3)}}_\eta(0,0) \sim 0.12 \textrm{ (GeV)}^{-1}$. For the $\eta'$ we use the same parameters as in Ref.~\cite{Chao:2020kwq}. For the total set of model parameters we have 
 $(m_{\pi},m_{V,\pi},f_\pi)=(345,916,101)$ MeV, $(m_{\eta},m_{V,\eta},f_\eta)=(468,900,122)$ MeV and $(m_{\eta'},m_{V,\eta'},f_{\eta'})=(982,952,74)$ MeV.

\section{Results}
\label{sect:results}
\begin{figure}
        \begin{subfigure}{0.49\textwidth}
            \centering
            \includegraphics[width=0.99\textwidth]{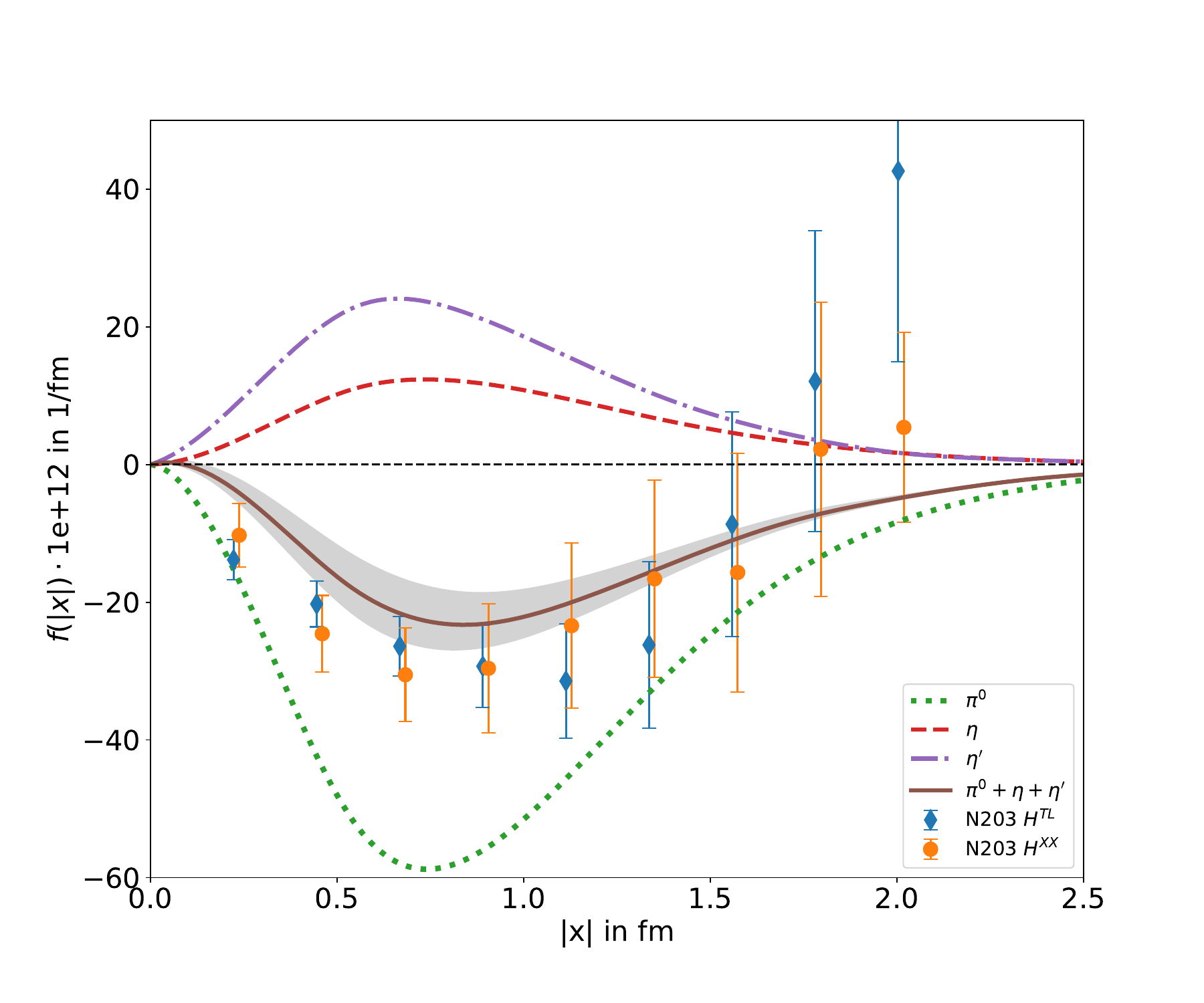}
            \subcaption{(2+2)a contribution to $a_\mu^{HVP,NLO}$}
            \label{fig:2+2a_a_mu}
        \end{subfigure}
        \begin{subfigure}{0.49\textwidth}
            \centering
            \includegraphics[width=0.99\textwidth]{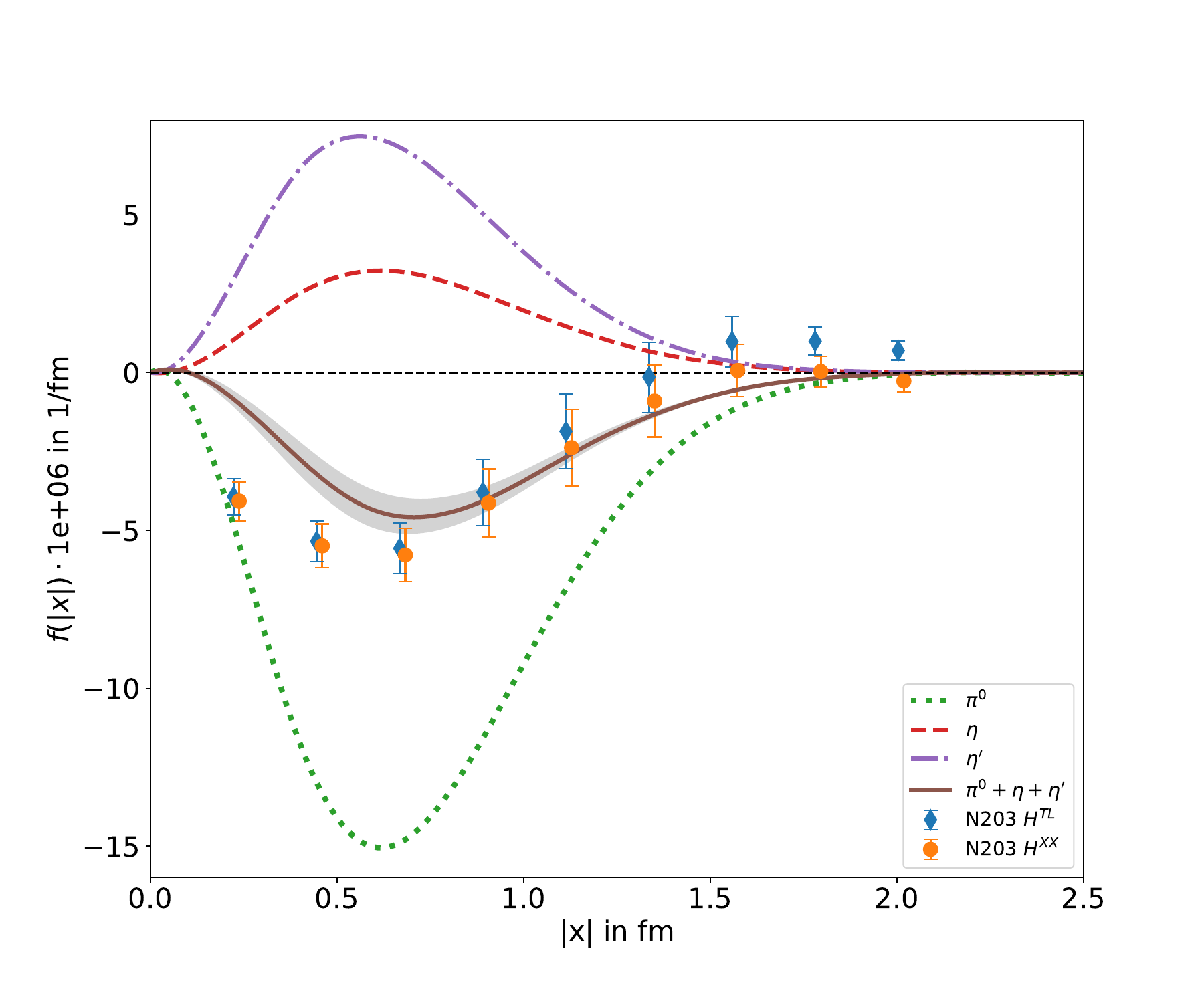}
            \subcaption{(2+2)a contribution to $\Pi(1GeV^2)-\Pi(0.25GeV^2)$}
            \label{fig:2+2a_vacpol}
        \end{subfigure}
        \caption{Comparison between the lattice data on one ensemble (N203) for the 'TL' and 'XX' kernel, defined in Eq.~\eref{eq:kernels}, and the pseudoscalar exchange model defined in Sect.~\eref{sect:pme}. The 'XX' kernel is slightly displaced for better readability.}
        \label{fig:2+2a_results}
\end{figure}

At first, we want to focus on  the $(2+2)a$ contribution. Since this contribution is UV-finite, we  drop the Pauli-Villars regulator. This also means that the continuum result for this quantity does not depend on the renormalization scheme and can be compared among different collaborations, similar to the leading-order HVP contribution. We display the integrand of Eq.~\eref{eq:2+2_master_a} for the lattice data using the 'TL' and 'XX' kernel, defined in Eq.~\eref{eq:kernels} together with the prediction of the pseudoscalar meson exchange in Fig.~\eref{fig:2+2a_a_mu}. We see a good agreement between the two kernel functions, providing a first check for our method. We also observe that the combined curve of the $\pi^0$, $\eta$ and $\eta'$ gives a semi-quantitative description of the lattice data.
By changing the weight functions, we can also investigate the subtracted vacuum polarization $\Pi(1GeV^2)-\Pi(0.25GeV^2)$. This quantity is much shorter-ranged, which has an improved signal quality. We see that the model describes the lattice data well for $|x|>0.7$ fm, however in the small-$|x|$ regime the model prediction differs from the lattice data.

We have to mention here that using the physical values for $m_{\eta'}$ and $f_{\eta'}$ of the $\eta'$ may overestimate its contribution for an ensemble which has a pion mass that is much larger than the physical one. The contributions of the $\pi^0$ and $\eta$ both get smaller in absolute size, when increasing the pion mass. For the $\eta'$ its dependence on the pion mass is not easy to predict, but it could follow a similar behaviour.
In order to improve the model, one would need to further investigate this. To get an approximation for the error of the model, we perform a fit of the model to the lattice data, where $f_{\eta'}$ is taken as a fit parameter. The uncertainty of the model is obtained from the error of this fit. We give the values for the integrated results in Table~\eref{tab:integrated}.
Using the model, we are able to make an estimate for the physical point. By choosing the PDG values \cite{ParticleDataGroup:2022pth} for the masses and two-photon couplings ${\cal F}(0,0)$ of the $\pi^0$, $\eta$ and $\eta'$ contribution we obtain the result $-54(8)\cdot 10^{-12}$ for $a_\mu^{(2+2)a,PME}$.

\begin{figure}
        \begin{subfigure}{0.49\textwidth}
            \centering
            \includegraphics[width=0.99\textwidth]{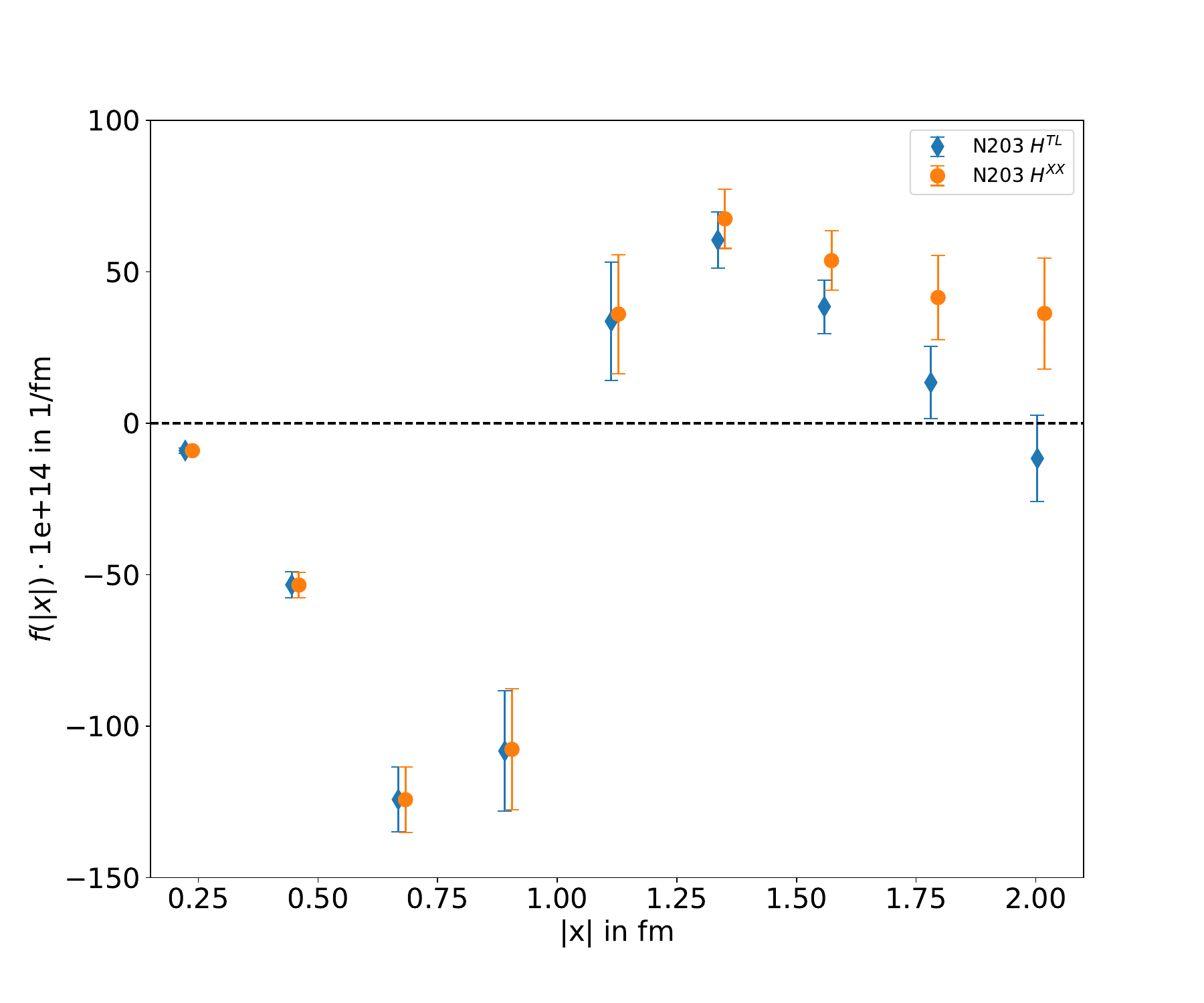}
            \subcaption{(2+2)a contribution to $a_\mu^{W}$}
            \label{fig:2+2b_a_mu}
        \end{subfigure}
        \begin{subfigure}{0.49\textwidth}
            \centering
            \includegraphics[width=0.99\textwidth]{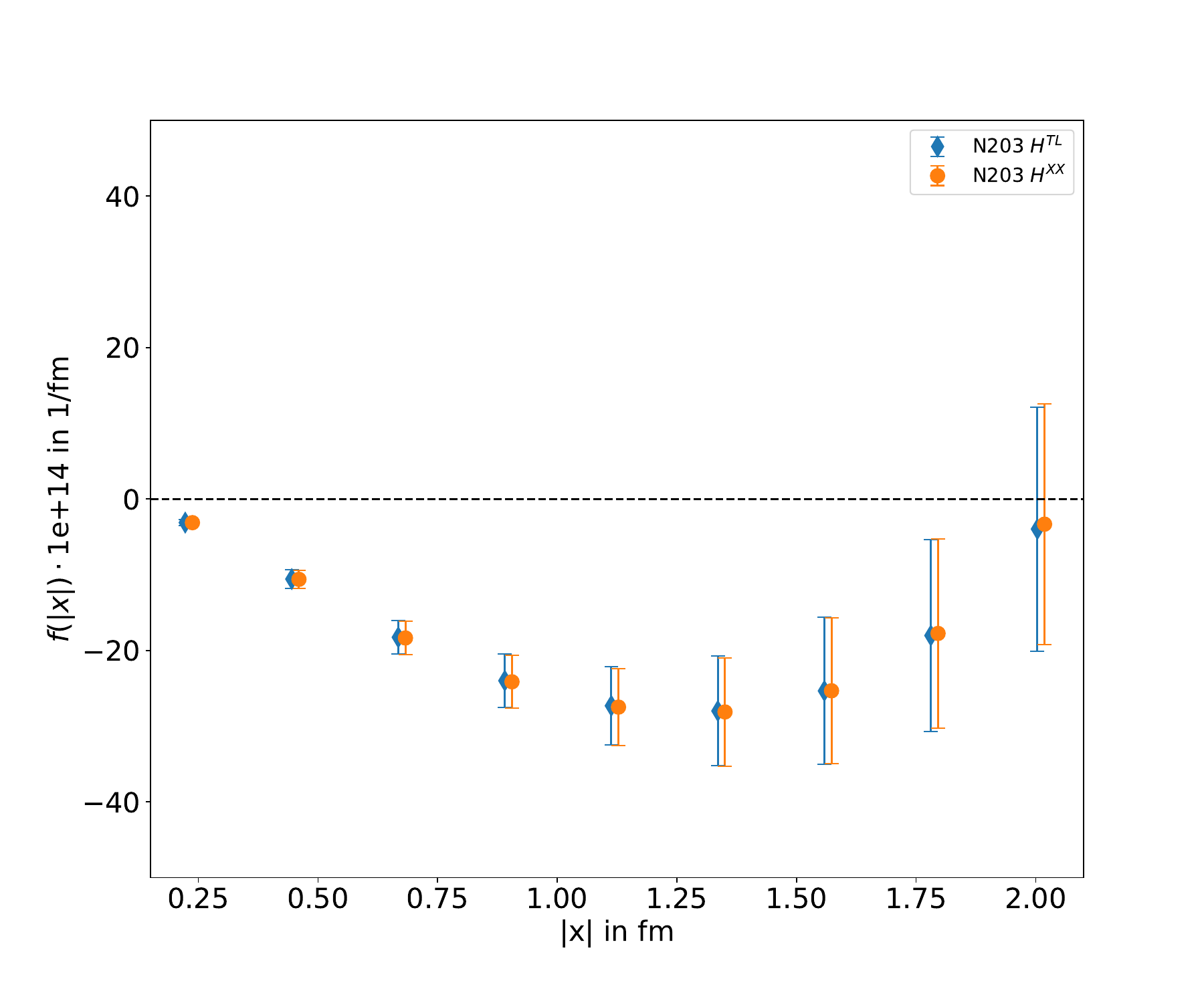}
            \subcaption{(2+2)b contribution to $a_\mu^{W}$}
            \label{fig:2+2b_vacpol}
        \end{subfigure}
        \caption{Comparison between the integrands for both (2+2) contributions to $a_\mu^W$ with a Pauli-Villars regulator \eref{eq:pv} with $\Lambda=3m_\mu$. The 'XX' kernel is slightly displaced for better readability.}
        \label{fig:2+2b_results}
\end{figure}

In contrast to the $(2+2)a$ contribution, the $(2+2)b$ contribution is UV-divergent. That means we can only obtain scheme-dependent results for this quantity. However, we can check the relative size of the $(2+2)a$ and $(2+2)b$ contribution with the same regulator with $\Lambda=3m_\mu$. We observe that the tail of the integrand of the $(2+2)b$ contribution to $a_\mu$ has not yet decayed to zero beyond $|x|>2$ fm, which is quite long-ranged. So, we chose to look at its contribution to the window quantity $a_\mu^W$ first, in order to have a better signal. 
The integrands for these contributions are shown in Fig.~\eref{fig:2+2b_results}. In contrast to the contributions displayed in Fig.~\eref{fig:2+2a_results}, the $(2+2)a$ contribution to $a_\mu^W$ shows a sign change at $|x|\sim 1$ fm, which reduces the total size of the integral. Comparing now the magnitude of both contributions we see that the $(2+2)b$ is of the same order as the $(2+2)a$. This means that for a full calculation of $a_\mu^{HVP,NLO}$ it is necessary to take this contribution into account.

\begin{table}
\footnotesize
    \centering
    \begin{tabular}{|c|c|c|c|c|} \hline  
         &  \multicolumn{3}{|c|}{$(2+2)a$}& $(2+2)b$\\ \hline  
 & $a_\mu \cdot 10^{12}$ & $[\Pi(1GeV^2)-\Pi(0.25GeV^2)]\cdot10^{7}$& $a_\mu^W \cdot 10^{14}$ $(\Lambda=3m_\mu)$&$a_\mu^W \cdot 10^{14}$ $(\Lambda=3m_\mu)$\\ \hline  
         $H^{TL}$&  -35(12)&  -43(13)&  -36(16)& -30(7)\\   
         $H^{XX}$&  -34(16)&  -50(14)&  -31(17)& -31(7)\\ \hline  
         $\pi^0$&  -68&  -124 \\   
         $\eta$&  15&  27\\  
         $\eta'$&  25&  59 \\ \cline{1-3}
         $\pi^0+\eta+\eta'$&  -28(4)&  -39(8) \\ \cline{1-3} 
    \end{tabular}
    
    \caption{Results for the integrated quantity. The results for the PME are integrated up to infinity. The lattice results for the subtracted vacuum polarization $\Pi(1GeV^2)-\Pi(0.25GeV^2)$ and the window quantity $a_\mu^W$ are integrated up to $|x|\sim 2$ fm, where the lattice data for $a_\mu^W$ is only integrated to $|x| \sim 1.8$ fm due to the large error of the data points at large $|x|$. The window quantity is calculated with a Pauli-Villars regulator for both $(2+2)$ contributions.}
    \label{tab:integrated}
\end{table}
\section{Conclusion}
We have shown that using the coordinate-space framework for the HVP at NLO $a_\mu^{HVP,NLO}$ proposed in Ref.~\cite{Biloshytskyi:2022ets} one is able to obtain results for the $(2+2)$ disconnected contributions to $a_\mu^{HVP,NLO}$ .
We compared the integrand of the UV-finite $(2+2)a$ contribution to the pseudoscalar meson exchange model with a VMD form factor, which is in good agreement with the lattice data on the ensemble with a pion mass of $m_\pi=346$ MeV. We then used the model to estimate this contribution at the physical point.
Since this contribution is UV-finite the result for this diagram does not depend on the renormalization scheme and it is in principle possible to isolate this contribution in a lattice calculation, to compare it among different collaborations as a crosscheck.\\
For the calculation of the UV-divergent $(2+2)b$ contribution it proves useful to apply the Pauli-Villars regularization scheme of the photon propagator proposed in Ref.~\cite{Biloshytskyi:2022ets}. We have seen that in this regularization scheme both $(2+2)$ contributions are equally important in the calculation of the window quantity. This suggests that in the calculation of $a_\mu^{HVP,NLO}$ it is also necessary to consider both contributions.
The proposed framework can also be used to calculate the fully-connected diagrams in order to obtain the dominant contribution to $a_\mu^{HVP,NLO}$ for fixed Pauli-Villars regulator. However, to obtain physical results, it will be necessary to include the counterterms and choose a renormalization scheme.

\vspace{0.3cm}
\footnotesize
{\textbf{Acknowledgements:}}
We acknowledge the support of Deutsche Forschungsgemeinschaft (DFG) through the research unit FOR~5327 ``Photon-photon interactions in the Standard Model and beyond exploiting the discovery potential from MESA to the LHC'' (grant 458854507), and through the Cluster of Excellence ``Precision Physics, Fundamental Interactions and Structure of Matter'' (PRISMA+ EXC 2118/1) funded within the German Excellence Strategy (project ID 39083149).
 E.-H.C.'s work was supported in part by the U.S. D.O.E. grant \#DE-SC0011941.
Calculations for this project were partly performed on the HPC clusters ``Clover'' and ``HIMster II'' at the Helmholtz-Institut Mainz and ``Mogon II'' at JGU Mainz. 
The measurement codes were developed based on the C++ library \texttt{wit}, an coding effort led by Renwick J.~Hudspith. 
We are grateful to our colleagues in the CLS initiative for sharing ensembles.

\bibliographystyle{apsrev4-1}
{\footnotesize
\bibliography{refs}

\begin{thebibliography}{15}%
\makeatletter
\providecommand \@ifxundefined [1]{%
 \@ifx{#1\undefined}
}%
\providecommand \@ifnum [1]{%
 \ifnum #1\expandafter \@firstoftwo
 \else \expandafter \@secondoftwo
 \fi
}%
\providecommand \@ifx [1]{%
 \ifx #1\expandafter \@firstoftwo
 \else \expandafter \@secondoftwo
 \fi
}%
\providecommand \natexlab [1]{#1}%
\providecommand \enquote  [1]{``#1''}%
\providecommand \bibnamefont  [1]{#1}%
\providecommand \bibfnamefont [1]{#1}%
\providecommand \citenamefont [1]{#1}%
\providecommand \href@noop [0]{\@secondoftwo}%
\providecommand \href [0]{\begingroup \@sanitize@url \@href}%
\providecommand \@href[1]{\@@startlink{#1}\@@href}%
\providecommand \@@href[1]{\endgroup#1\@@endlink}%
\providecommand \@sanitize@url [0]{\catcode `\\12\catcode `\$12\catcode `\&12\catcode `\#12\catcode `\^12\catcode `\_12\catcode `\%12\relax}%
\providecommand \@@startlink[1]{}%
\providecommand \@@endlink[0]{}%
\providecommand \url  [0]{\begingroup\@sanitize@url \@url }%
\providecommand \@url [1]{\endgroup\@href {#1}{\urlprefix }}%
\providecommand \urlprefix  [0]{URL }%
\providecommand \Eprint [0]{\href }%
\providecommand \doibase [0]{http://dx.doi.org/}%
\providecommand \selectlanguage [0]{\@gobble}%
\providecommand \bibinfo  [0]{\@secondoftwo}%
\providecommand \bibfield  [0]{\@secondoftwo}%
\providecommand \translation [1]{[#1]}%
\providecommand \BibitemOpen [0]{}%
\providecommand \bibitemStop [0]{}%
\providecommand \bibitemNoStop [0]{.\EOS\space}%
\providecommand \EOS [0]{\spacefactor3000\relax}%
\providecommand \BibitemShut  [1]{\csname bibitem#1\endcsname}%
\let\auto@bib@innerbib\@empty
\bibitem [{\citenamefont {Aguillard}\ \emph {et~al.}(2023)\citenamefont {Aguillard} \emph {et~al.}}]{Muong-2:2023cdq}%
  \BibitemOpen
  \bibfield  {author} {\bibinfo {author} {\bibfnamefont {D.~P.}\ \bibnamefont {Aguillard}} \emph {et~al.} (\bibinfo {collaboration} {Muon g-2}),\ }\href@noop {} {\  (\bibinfo {year} {2023})},\ \Eprint {http://arxiv.org/abs/2308.06230} {arXiv:2308.06230 [hep-ex]} \BibitemShut {NoStop}%
\bibitem [{\citenamefont {Aoyama}\ \emph {et~al.}(2020)\citenamefont {Aoyama} \emph {et~al.}}]{Aoyama:2020ynm}%
  \BibitemOpen
  \bibfield  {author} {\bibinfo {author} {\bibfnamefont {T.}~\bibnamefont {Aoyama}} \emph {et~al.},\ }\href {\doibase 10.1016/j.physrep.2020.07.006} {\bibfield  {journal} {\bibinfo  {journal} {Phys. Rept.}\ }\textbf {\bibinfo {volume} {887}},\ \bibinfo {pages} {1} (\bibinfo {year} {2020})},\ \Eprint {http://arxiv.org/abs/2006.04822} {arXiv:2006.04822 [hep-ph]} \BibitemShut {NoStop}%
\bibitem [{\citenamefont {Ignatov}\ \emph {et~al.}(2023)\citenamefont {Ignatov} \emph {et~al.}}]{CMD-3:2023alj}%
  \BibitemOpen
  \bibfield  {author} {\bibinfo {author} {\bibfnamefont {F.~V.}\ \bibnamefont {Ignatov}} \emph {et~al.} (\bibinfo {collaboration} {CMD-3}),\ }\href@noop {} {\  (\bibinfo {year} {2023})},\ \Eprint {http://arxiv.org/abs/2302.08834} {arXiv:2302.08834 [hep-ex]} \BibitemShut {NoStop}%
\bibitem [{\citenamefont {Borsanyi}\ \emph {et~al.}(2021)\citenamefont {Borsanyi} \emph {et~al.}}]{Borsanyi:2020mff}%
  \BibitemOpen
  \bibfield  {author} {\bibinfo {author} {\bibfnamefont {S.}~\bibnamefont {Borsanyi}} \emph {et~al.},\ }\href {\doibase 10.1038/s41586-021-03418-1} {\bibfield  {journal} {\bibinfo  {journal} {Nature}\ }\textbf {\bibinfo {volume} {593}},\ \bibinfo {pages} {51} (\bibinfo {year} {2021})},\ \Eprint {http://arxiv.org/abs/2002.12347} {arXiv:2002.12347 [hep-lat]} \BibitemShut {NoStop}%
\bibitem [{\citenamefont {Chao}\ \emph {et~al.}(2021)\citenamefont {Chao}, \citenamefont {Hudspith}, \citenamefont {G\'erardin}, \citenamefont {Green}, \citenamefont {Meyer},\ and\ \citenamefont {Ottnad}}]{Chao:2021tvp}%
  \BibitemOpen
  \bibfield  {author} {\bibinfo {author} {\bibfnamefont {E.-H.}\ \bibnamefont {Chao}}, \bibinfo {author} {\bibfnamefont {R.~J.}\ \bibnamefont {Hudspith}}, \bibinfo {author} {\bibfnamefont {A.}~\bibnamefont {G\'erardin}}, \bibinfo {author} {\bibfnamefont {J.~R.}\ \bibnamefont {Green}}, \bibinfo {author} {\bibfnamefont {H.~B.}\ \bibnamefont {Meyer}}, \ and\ \bibinfo {author} {\bibfnamefont {K.}~\bibnamefont {Ottnad}},\ }\href {\doibase 10.1140/epjc/s10052-021-09455-4} {\bibfield  {journal} {\bibinfo  {journal} {Eur. Phys. J. C}\ }\textbf {\bibinfo {volume} {81}},\ \bibinfo {pages} {651} (\bibinfo {year} {2021})},\ \Eprint {http://arxiv.org/abs/2104.02632} {arXiv:2104.02632 [hep-lat]} \BibitemShut {NoStop}%
\bibitem [{\citenamefont {Blum}\ \emph {et~al.}(2023)\citenamefont {Blum}, \citenamefont {Christ}, \citenamefont {Hayakawa}, \citenamefont {Izubuchi}, \citenamefont {Jin}, \citenamefont {Jung}, \citenamefont {Lehner},\ and\ \citenamefont {Tu}}]{Blum:2023vlm}%
  \BibitemOpen
  \bibfield  {author} {\bibinfo {author} {\bibfnamefont {T.}~\bibnamefont {Blum}}, \bibinfo {author} {\bibfnamefont {N.}~\bibnamefont {Christ}}, \bibinfo {author} {\bibfnamefont {M.}~\bibnamefont {Hayakawa}}, \bibinfo {author} {\bibfnamefont {T.}~\bibnamefont {Izubuchi}}, \bibinfo {author} {\bibfnamefont {L.}~\bibnamefont {Jin}}, \bibinfo {author} {\bibfnamefont {C.}~\bibnamefont {Jung}}, \bibinfo {author} {\bibfnamefont {C.}~\bibnamefont {Lehner}}, \ and\ \bibinfo {author} {\bibfnamefont {C.}~\bibnamefont {Tu}},\ }\href@noop {} {\  (\bibinfo {year} {2023})},\ \Eprint {http://arxiv.org/abs/2304.04423} {arXiv:2304.04423 [hep-lat]} \BibitemShut {NoStop}%
\bibitem [{\citenamefont {Meyer}(2017)}]{Meyer:2017hjv}%
  \BibitemOpen
  \bibfield  {author} {\bibinfo {author} {\bibfnamefont {H.~B.}\ \bibnamefont {Meyer}},\ }\href {\doibase 10.1140/epjc/s10052-017-5200-3} {\bibfield  {journal} {\bibinfo  {journal} {Eur. Phys. J. C}\ }\textbf {\bibinfo {volume} {77}},\ \bibinfo {pages} {616} (\bibinfo {year} {2017})},\ \Eprint {http://arxiv.org/abs/1706.01139} {arXiv:1706.01139 [hep-lat]} \BibitemShut {NoStop}%
\bibitem [{\citenamefont {Chao}\ \emph {et~al.}(2023)\citenamefont {Chao}, \citenamefont {Meyer},\ and\ \citenamefont {Parrino}}]{Chao:2022ycy}%
  \BibitemOpen
  \bibfield  {author} {\bibinfo {author} {\bibfnamefont {E.-H.}\ \bibnamefont {Chao}}, \bibinfo {author} {\bibfnamefont {H.~B.}\ \bibnamefont {Meyer}}, \ and\ \bibinfo {author} {\bibfnamefont {J.}~\bibnamefont {Parrino}},\ }\href {\doibase 10.1103/PhysRevD.107.054505} {\bibfield  {journal} {\bibinfo  {journal} {Phys. Rev. D}\ }\textbf {\bibinfo {volume} {107}},\ \bibinfo {pages} {054505} (\bibinfo {year} {2023})},\ \Eprint {http://arxiv.org/abs/2211.15581} {arXiv:2211.15581 [hep-lat]} \BibitemShut {NoStop}%
\bibitem [{\citenamefont {Biloshytskyi}\ \emph {et~al.}(2023)\citenamefont {Biloshytskyi}, \citenamefont {Chao}, \citenamefont {G\'erardin}, \citenamefont {Green}, \citenamefont {Hagelstein}, \citenamefont {Meyer}, \citenamefont {Parrino},\ and\ \citenamefont {Pascalutsa}}]{Biloshytskyi:2022ets}%
  \BibitemOpen
  \bibfield  {author} {\bibinfo {author} {\bibfnamefont {V.}~\bibnamefont {Biloshytskyi}}, \bibinfo {author} {\bibfnamefont {E.-H.}\ \bibnamefont {Chao}}, \bibinfo {author} {\bibfnamefont {A.}~\bibnamefont {G\'erardin}}, \bibinfo {author} {\bibfnamefont {J.~R.}\ \bibnamefont {Green}}, \bibinfo {author} {\bibfnamefont {F.}~\bibnamefont {Hagelstein}}, \bibinfo {author} {\bibfnamefont {H.~B.}\ \bibnamefont {Meyer}}, \bibinfo {author} {\bibfnamefont {J.}~\bibnamefont {Parrino}}, \ and\ \bibinfo {author} {\bibfnamefont {V.}~\bibnamefont {Pascalutsa}},\ }\href {\doibase 10.1007/JHEP03(2023)194} {\bibfield  {journal} {\bibinfo  {journal} {JHEP}\ }\textbf {\bibinfo {volume} {03}},\ \bibinfo {pages} {194} (\bibinfo {year} {2023})},\ \Eprint {http://arxiv.org/abs/2209.02149} {arXiv:2209.02149 [hep-lat]} \BibitemShut {NoStop}%
\bibitem [{\citenamefont {Bruno}\ \emph {et~al.}(2017)\citenamefont {Bruno}, \citenamefont {Korzec},\ and\ \citenamefont {Schaefer}}]{Bruno:2016plf}%
  \BibitemOpen
  \bibfield  {author} {\bibinfo {author} {\bibfnamefont {M.}~\bibnamefont {Bruno}}, \bibinfo {author} {\bibfnamefont {T.}~\bibnamefont {Korzec}}, \ and\ \bibinfo {author} {\bibfnamefont {S.}~\bibnamefont {Schaefer}},\ }\href {\doibase 10.1103/PhysRevD.95.074504} {\bibfield  {journal} {\bibinfo  {journal} {Phys. Rev. D}\ }\textbf {\bibinfo {volume} {95}},\ \bibinfo {pages} {074504} (\bibinfo {year} {2017})},\ \Eprint {http://arxiv.org/abs/1608.08900} {arXiv:1608.08900 [hep-lat]} \BibitemShut {NoStop}%
\bibitem [{\citenamefont {C\`e}\ \emph {et~al.}(2022)\citenamefont {C\`e} \emph {et~al.}}]{Ce:2022kxy}%
  \BibitemOpen
  \bibfield  {author} {\bibinfo {author} {\bibfnamefont {M.}~\bibnamefont {C\`e}} \emph {et~al.},\ }\href {\doibase 10.1103/PhysRevD.106.114502} {\bibfield  {journal} {\bibinfo  {journal} {Phys. Rev. D}\ }\textbf {\bibinfo {volume} {106}},\ \bibinfo {pages} {114502} (\bibinfo {year} {2022})},\ \Eprint {http://arxiv.org/abs/2206.06582} {arXiv:2206.06582 [hep-lat]} \BibitemShut {NoStop}%
\bibitem [{\citenamefont {Chao}\ \emph {et~al.}(2020)\citenamefont {Chao}, \citenamefont {G\'erardin}, \citenamefont {Green}, \citenamefont {Hudspith},\ and\ \citenamefont {Meyer}}]{Chao:2020kwq}%
  \BibitemOpen
  \bibfield  {author} {\bibinfo {author} {\bibfnamefont {E.-H.}\ \bibnamefont {Chao}}, \bibinfo {author} {\bibfnamefont {A.}~\bibnamefont {G\'erardin}}, \bibinfo {author} {\bibfnamefont {J.~R.}\ \bibnamefont {Green}}, \bibinfo {author} {\bibfnamefont {R.~J.}\ \bibnamefont {Hudspith}}, \ and\ \bibinfo {author} {\bibfnamefont {H.~B.}\ \bibnamefont {Meyer}},\ }\href {\doibase 10.1140/epjc/s10052-020-08444-3} {\bibfield  {journal} {\bibinfo  {journal} {Eur. Phys. J. C}\ }\textbf {\bibinfo {volume} {80}},\ \bibinfo {pages} {869} (\bibinfo {year} {2020})},\ \Eprint {http://arxiv.org/abs/2006.16224} {arXiv:2006.16224 [hep-lat]} \BibitemShut {NoStop}%
\bibitem [{\citenamefont {Knecht}\ and\ \citenamefont {Nyffeler}(2002)}]{Knecht:2001qf}%
  \BibitemOpen
  \bibfield  {author} {\bibinfo {author} {\bibfnamefont {M.}~\bibnamefont {Knecht}}\ and\ \bibinfo {author} {\bibfnamefont {A.}~\bibnamefont {Nyffeler}},\ }\href {\doibase 10.1103/PhysRevD.65.073034} {\bibfield  {journal} {\bibinfo  {journal} {Phys. Rev. D}\ }\textbf {\bibinfo {volume} {65}},\ \bibinfo {pages} {073034} (\bibinfo {year} {2002})},\ \Eprint {http://arxiv.org/abs/hep-ph/0111058} {arXiv:hep-ph/0111058} \BibitemShut {NoStop}%
\bibitem [{\citenamefont {G\'erardin}\ \emph {et~al.}(2019)\citenamefont {G\'erardin}, \citenamefont {C\`e}, \citenamefont {von Hippel}, \citenamefont {H{\"o}rz}, \citenamefont {Meyer}, \citenamefont {Mohler}, \citenamefont {Ottnad}, \citenamefont {Wilhelm},\ and\ \citenamefont {Wittig}}]{gerardin:2019rua}%
  \BibitemOpen
  \bibfield  {author} {\bibinfo {author} {\bibfnamefont {A.}~\bibnamefont {G\'erardin}}, \bibinfo {author} {\bibfnamefont {M.}~\bibnamefont {C\`e}}, \bibinfo {author} {\bibfnamefont {G.}~\bibnamefont {von Hippel}}, \bibinfo {author} {\bibfnamefont {B.}~\bibnamefont {H{\"o}rz}}, \bibinfo {author} {\bibfnamefont {H.~B.}\ \bibnamefont {Meyer}}, \bibinfo {author} {\bibfnamefont {D.}~\bibnamefont {Mohler}}, \bibinfo {author} {\bibfnamefont {K.}~\bibnamefont {Ottnad}}, \bibinfo {author} {\bibfnamefont {J.}~\bibnamefont {Wilhelm}}, \ and\ \bibinfo {author} {\bibfnamefont {H.}~\bibnamefont {Wittig}},\ }\href {\doibase 10.1103/PhysRevD.100.014510} {\bibfield  {journal} {\bibinfo  {journal} {Phys. Rev.}\ }\textbf {\bibinfo {volume} {D100}},\ \bibinfo {pages} {014510} (\bibinfo {year} {2019})},\ \Eprint {http://arxiv.org/abs/1904.03120} {arXiv:1904.03120 [hep-lat]} \BibitemShut {NoStop}%
\bibitem [{\citenamefont {Workman}\ \emph {et~al.}(2022)\citenamefont {Workman} \emph {et~al.}}]{ParticleDataGroup:2022pth}%
  \BibitemOpen
  \bibfield  {author} {\bibinfo {author} {\bibfnamefont {R.~L.}\ \bibnamefont {Workman}} \emph {et~al.} (\bibinfo {collaboration} {Particle Data Group}),\ }\href {\doibase 10.1093/ptep/ptac097} {\bibfield  {journal} {\bibinfo  {journal} {PTEP}\ }\textbf {\bibinfo {volume} {2022}},\ \bibinfo {pages} {083C01} (\bibinfo {year} {2022})}\BibitemShut {NoStop}%
\end{thebibliography}%
}

\end{document}